\documentclass[sigconf]{acmart}

\AtBeginDocument{%
  }

\setcopyright{rightsretained}
\copyrightyear{2024}
\acmYear{2024}
\acmConference[LOCO '24]{LOCO '24: 1st International Workshop on Low Carbon Computing}{December 03, 2024}{Glasgow, UK/Online}
\acmBooktitle{LOCO '24: 1st International Workshop on Low Carbon Computing, December 03, 2024}

\begin{document}

\title{Assessing the Ecological Impact of AI}

\author{Sylvia Wenmackers}
\email{sylvia.wenmackers@kuleuven.be}
\orcid{0000-0002-1041-3533}
\affiliation{%
  \institution{Centre for Logic and Philosophy of Science (CLPS)}
  \city{KU Leuven}
  \country{Belgium}
}

\begin{abstract}
Philosophers of technology have recently started paying more attention to the environmental impacts of AI, in particular of large language models (LLMs) and generative AI (genAI) applications.
Meanwhile, few developers of AI give concrete estimates of the ecological impact of their models and products, and even when they do so, their analysis is often limited to green house gas emissions of certain stages of AI development or use.
The current proposal encourages practically viable analyses of the sustainability aspects of genAI informed by philosophical ideas.
\end{abstract}


\maketitle

\section*{Form Ethics of Technology to Engineering \ldots}
Bolte \& van Wynsberghe (2024) \cite{BoltevanWynsberghe:2025} signaled the rise of `Sustainable AI' as a new subfield of AI ethics, an evolution which van Wynsberghe (2021) \cite{vanWynsberghe:2021} called the `third wave of AI ethics' and which the authors consider to be part of the `Terrestrial Turn in the philosophy of technology' as Lemmens et al.\ (2017) called it in an editorial \cite{Lemmens-etal:2017}. Previously, authors such as Bostrom \& Yudkowsky (2018) \cite{BostromYudkowsky:2018} dealt primarily with long-term existential risks associated with speculative future AI developments, while Bender et al.\ (2021) \cite[§6.2]{Bender-etal:2024} and others focus on case studies on the concrete harms by existing and forthcoming AI models (due to algorithmic bias, for instance).

Philosophers writing on Sustainable AI flag the need for a critical analysis of the (un-)sustainability of AI itself in contrast to focusing on the potential of applications of AI for, say, fighting climate change. The latter may be linked to green-washing, which Heilinger et al.\ (2024) defined as: ``the particular form of ethics-washing that focuses on and attempts to obscure the environmental impact of a given AI technology'' \cite[§5]{Heilinger-etal:2024}.

The literature on Sustainable AI is clearly motivated by pressing ecological and societal needs, so it seems especially relevant to ask how the findings of AI ethicists can be translated into insights that inform engineers and other researchers, who develop and/or apply AI, as well as policy makers, who prepare decisions on AI investments. This raises the following question: \textbf{how do we make the ethical discussion on Sustainable AI actionable for engineers and policy makers?}

The ecological sustainability of new products is often assessed with a life cycle assessment (LCA), which considers the potential ecological impacts from the energy and raw materials needed to create the product (cradle), to use it, and to dispose of it (grave). Currently, many researchers use genAI in their work and engineers include it in many applications. So, it seems vital to estimate the impact of AI in the LCA as well. Likewise, the impact of genAI on socio-economic aspects of sustainability should also be assessed, for instance in a product line analysis (PLA), but starting with an LCA is a crucial first step in that direction.

So, \textbf{my concrete proposal is to start from LCAs as a methodolgy that is already in use by R\&D departments and to expand it to include the ecological impact of AI}.

Unfortunately, there is much more literature on using genAI to draw up LCAs than there are examples of sustainability assessments of AI applications. Still, there are a few extant LCAs or similar assessments of genAI models:
\begin{itemize}
    \item Berthelot et al.\ (2024) presented an ``LCA-based methodology for a multi-criteria evaluation of the environmental impact of generative AI services'' applied to one genAI image generator (Stable Diffusion) \cite{Berthelot-etal:2024}.
    \item Luccioni et al.\ (2023) estimated the carbon footprint of an open-science language model with 176 billion parameters (BLOOM); although they lacked some ``necessary information to carry out a `cradle-to-grave' assessment'', they aspired to an LCA-like analysis \cite{Luccioni-etal:2023}.
    \item Delort et al.\ (2023) reviewed much of the relevant literature, with further examples of (partial) assessments \cite{Delort-etal:2023}.
\end{itemize}
These sources, which focus on genAI models as such, may serve as a crucial starting point for assessing the sustainability of products that apply these models.

While concrete workflows and methodologies (such as LCA) are needed to give ethical considerations real-world impact, sustained contact with the ethics-of-AI community may foster a critical attitude to avoid overly shallow implementations. Just like healthy food is not a property of an isolated food item eaten by an individual at a specific occasion, also sustainability should be understood ``as a property of complex systems'' rather than of AI models in isolation (Bolte et al., 2022 \cite[§6]{Bolte-etal:2022}). Current environmental analyses, however, if they are made at all, tend to have a narrow focus, often on the carbon footprint of a single AI model (see, e.g., Simon et al., 2024 \cite{Simon-etal:2025}). They also favour the direct impact, over higher-order effects, including behavioral change (Hilty \& Hercheui, 2010 \cite{HiltyHercheui:2010}; but see Bieser, 2024 \cite{Bieser:2024}). Like in the debate on short- versus long-term AI risks, it seems needed to make a full-scale analysis that includes highly likely, direct impacts as well as more speculative, indirect ones (cf.\ S{\ae}tra \& Danaher, 2023 \cite{SaetraDanaher:2023}).

\section*{\ldots and back}
Vice versa, the proposed dialogue between philosophers of technology and engineers may also broaden the scope of the former. For instance, genAI is not the only technology that raises urgent environmental concerns: similar worries apply to cryptocurrencies and the internet of things (IoT) (see, e.g., Freitag et al., 2021 \cite{Freitag-etal:2021}).

Moreover, fostering interdisciplinary discussions may maximize the uptake in policy documents, as Hu et al.\ (2024) showed for research on the COVID-19 pandemic \cite{Hu-etal:2024}.

\bibliographystyle{ACM-Reference-Format}
\bibliography{references}

\end{document}